\documentclass[10pt,english,aip,apl,twocolumn]{revtex4-1}
\usepackage[T1]{fontenc}
\usepackage[latin9]{inputenc}
\usepackage{units}
\usepackage{amstext}
\usepackage{graphicx}

\makeatletter

\@ifundefined{textcolor}{}
{%
 \definecolor{BLACK}{gray}{0}
 \definecolor{WHITE}{gray}{1}
 \definecolor{RED}{rgb}{1,0,0}
 \definecolor{GREEN}{rgb}{0,1,0}
 \definecolor{BLUE}{rgb}{0,0,1}
 \definecolor{CYAN}{cmyk}{1,0,0,0}
 \definecolor{MAGENTA}{cmyk}{0,1,0,0}
 \definecolor{YELLOW}{cmyk}{0,0,1,0}
}

\makeatother

\usepackage{babel}
\begin{document}

\title{Thermal Creation of Electron Spin Polarization in n-Type Silicon}

\author{André Dankert}

\email{andre.dankert@chalmers.se}

\selectlanguage{english}%

\affiliation{Department of Microtechnology and Nanoscience, Chalmers University
of Technology, SE-41296, Göteborg, Sweden}

\author{Saroj P. Dash}

\email{saroj.dash@chalmers.se}

\selectlanguage{english}%

\affiliation{Department of Microtechnology and Nanoscience, Chalmers University
of Technology, SE-41296, Göteborg, Sweden}
\begin{abstract}
Conversion of heat into a spin-current in electron doped silicon can
offer a promising path for spin-caloritronics. Here we create an electron
spin polarization in the conduction band of n-type silicon by producing
a temperature gradient across a ferromagnetic tunnel contact. The
substrate heating experiments induce a large spin signal of $\unit[95]{\mu V}$,
corresponding to $\unit[0.54]{meV}$ spin-splitting in the conduction
band of n-type silicon by Seebeck spin tunneling mechanism. The thermal
origin of the spin injection has been confirmed by the quadratic scaling
of the spin signal with the Joule heating current and linear dependence
with the heating power. 
\end{abstract}


\maketitle
Ultra large scale integration and miniaturization of electronic devices
and circuitry yield to an increase in Joule heat dissipation and wastage
of energy. An effort to reduce, reuse or redirect this heat is not
only of technological, but also has fundamental and ecological importance.
In this regard caloritronics explores the possibility of directly
converting heat into electricity \cite{Snyder2008}. On the other
hand spintronics offers a promising path to achieve lower energy consumption
\cite{Zutic2004,Fert2008}. An exciting new research field taking
advantage of both of these areas, called spin-caloritronics, is attracting
a growing interest \cite{Bauer2012,Johnson1987a}. Recently, many
physical phenomena connecting spin to thermal effects have been proposed
and demonstrated such as spin-Seebeck effect \cite{Uchida2008}, spin
dependent Seebeck effect in nanostructures and thermal spin injection
into metal \cite{Slachter2010}, thermal spin-transfer torque \cite{Hatami2007,Yu2010},
spin-related Peltier cooling \cite{Flipse2012}, and thermal spin
tunnel effects \cite{Walter2011,LeBreton2011a}. Especially interesting
is the Seebeck spin tunneling into semiconductors, which is sensitive
to the energy derivative of tunnel spin polarization at the interfaces
\cite{LeBreton2011a}.

The strong interest in Si for spintronics stems from its long spin
coherence length, caused by the absence of hyperfine interactions
and a weak spin-orbit coupling, and its industrial dominance \cite{Jansen2012c,Jansen2012a}.
Recently, electrical spin polarization in Si could be created at room
temperature by spin polarized tunneling from a ferromagnet (FM) \cite{Dash2009,Jansen2010,Sasaki2009a,Jeon2011,Li2011}.
The Hanle and the inverse of the Hanle effects were used for the detection
of spin polarization in different semiconductors under ferromagnetic
tunnel contacts \cite{Dash2011a}. Using such Hanle techniques, Le
Breton et al. demonstrated a spin accumulation in hole doped p-type
Si by using the Seebeck spin tunneling mechanism \cite{LeBreton2011a}.
In a FM/tunnel barrier/semiconductor structure, a temperature difference
between the electrodes creates a tunnel spin current \cite{LeBreton2011a,Jansen2012,Jain2012,Jeon2012a},
which is governed by the energy derivative of the tunnel spin polarization
(TSP). This involves thermal transfer of spin angular momentum from
the ferromagnet to Si without net tunneling charge current. For Si
based devices, these experiments are yet limited to reports on hole
doped p-type Si with Al$_{2}$O$_{3}$ / NiFe tunnel contacts \cite{LeBreton2011a}.
For further development, it is required to have spin tunnel contacts
with strong variation in TSP around the Fermi energy (EF), tunnel
barriers with high interface thermal resistances, and its implementation
in both electron and hole doped Si \cite{LeBreton2011a}. Therefore,
a detailed investigation of spin Seebeck tunneling effect with new
ferromagnetic tunnel contact materials, and electron doped n-type
Si is crucial from fundamental and technological point of view.

\begin{figure}
\begin{centering}
\includegraphics{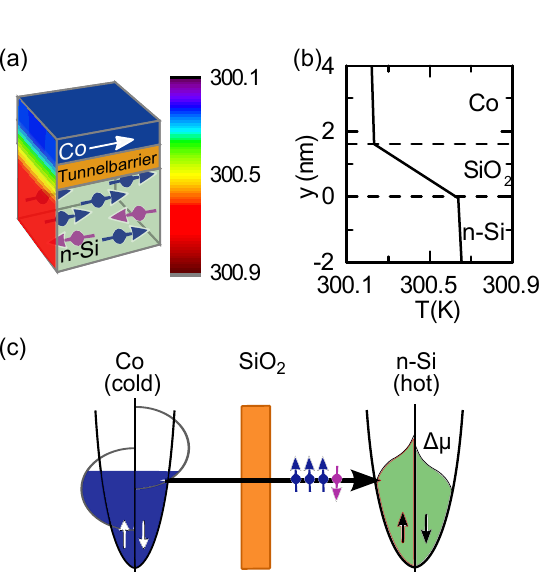} 
\par\end{centering}

\protect\caption{\textbf{(Color online) }(a) Device structure and simulated temperature
distribution across the Co/SiO$_{2}$/n-type Si. (b) Temperature line
scan across the tunnel contact for $\unit[1.5]{nm}$ thick SiO$_{2}$
at heating current $\unit[20]{kAcm^{-2}}$. (c) Spin-dependent density
of states and its occupation for a tunnel contact with a hot Si and
a cold Co electrode. The different energy dependence of tunnel spin
polarization for forward and reverse tunneling process gives rise
to a finite spin current, without any net charge current.}

\label{fig1} 
\end{figure}

In this letter, we report a thermal creation of electron spin polarization
in n-type Si by using the Seebeck spin tunneling mechanism. The temperature
gradient between the n-type Si substrate and the FM yields to Seebeck
spin tunnel currents, creating a large spin accumulation in the Si
conduction band. We particularly use ferromagnetic tunnel contacts
with ozone oxidized SiO$_{2}$ tunnel barriers and Co ferromagnet
on heavily doped n-type Si. The detection of this thermally induced
spin polarization was performed by employing Hanle and inverted Hanle
effects in four-terminal measurement geometry. The amplitude of the
thermal spin signal scales quadratically with the Joule heating current
density and linearly with the heating power, which confirms the thermal
origin of the observed Hanle signal. Electrical measurements and high
magnetic field experiments rule out any other type of thermomagnetic
effects. This result offers a way to generate a large electron spin
accumulation in n-type Si by thermal spin injection using Seebeck
spin tunneling.

For thermal spin injection into the n-type Si (measured electron density
of $\unit[3\cdot10^{19}]{cm^{-3}}$ at $\unit[300]{K}$) FM tunnel
contacts (Co/SiO$_{2}$) were fabricated on initially patterned $\unit[2]{\mu m}$
thick silicon-on-insulator (SOI) channels (with the dimensions $\unit[200\times1000]{\mu m^{2}}$).
The SiO$_{2}$ tunnel barrier was formed on the Si by ozone oxidation
at room temperature. The chips were exposed to O$_{3}$, created by
a constant O$_{2}$ flow ($\unit[1]{l/min}$) in the presence of a
UV radiation source. This process results in uniform and pinhole-free
tunnel barriers of $\unit[1.5]{nm}$ thickness. This is confirmed
by the temperature and thickness dependent measurements of the junction
resistance \cite{Dankert2013}. Using such SiO$_{2}$ tunnel barrier,
a very large spin polarization in silicon could be achieved by electrical
method \cite{Dankert2013}. After oxidation the samples were transferred
to an electron beam deposition system, where $\unit[15]{nm}$ Co and
10 nm Au were deposited. The FM contacts were patterned by photolithography
and Ar\textendash ion beam etching. Subsequently, Cr ($\unit[10]{nm}$)/Au
($\unit[100]{nm}$) contacts on the Si, and contact pads on the ferromagnetic
tunnel contacts were prepared by photolithography and lift-off technique.
The Cr/Au contacts were used for the Joule heating of the Si and as
reference contacts to detect a voltage signal with respect to the
FM tunnel contact. The ferromagnetic tunnel contacts have an area
of $\unit[200\times100]{\mu m^{2}}$ allowing a sufficiently good
signal-to-noise ratio.

\begin{figure}[h]
\begin{centering}
\includegraphics[scale=0.97]{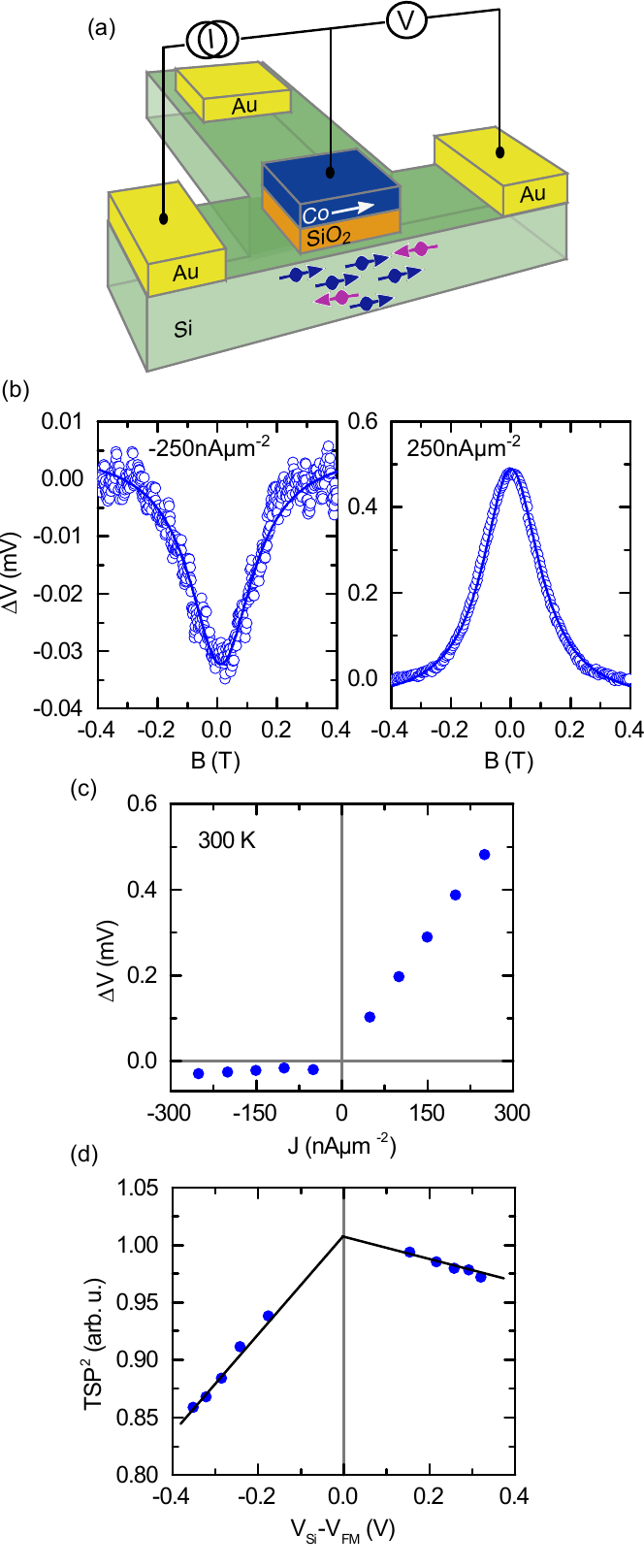} 
\par\end{centering}

\protect\caption{\textbf{(Color online) Electrical Hanle effect at room temperature.}
(a) Three-terminal Hanle geometry for electrical injection and detection
of spin polarization in the n-type Si with a SiO$_{2}$-Co tunnel
contact. (b) The electrical detection of spin polarizations in n-type
Si at $\unit[300]{K}$ by Hanle effect for spin injection ($I=\unit[+250]{nA\mu m^{-2}}$)
and spin extraction ($I=\unit[-250]{nA\mu m^{-2}}$). The solid lines
are the Lorentzian fits with lower limit for spin life time $\tau_{sf}=\unit[50]{ps}$.
(c) Electrical bias current dependence of the Hanle voltage signal
($\Delta V$). (d) Electrical bias voltage dependence of $TSP^{2}$,
which is proportional to the Spin-RA of the detected Hanle signal.}

\label{fig2} 
\end{figure}

The principle for the thermal creation of spin accumulation in Si
using the Seebeck spin tunneling mechanism is illustrated in Fig.
\ref{fig1}. The simulated temperature distribution across the interface
of the Co/SiO$_{2}$ ($\unit[1.5]{nm}$)/n-type Si device is shown
in Fig. 1a and b. The complete device structure of Au ($\unit[100]{nm}$)/Co
($\unit[15]{nm}$)/SiO$_{2}$ ($\unit[1.5]{nm}$)/n-type Si ($\unit[2]{\mu m}$)
were considered for our calculation. A model was drawn to scale with
our device and divided into finite elements using commercial software
(COMSOL) for calculating their temperatures. The temperature gradient
($\Delta T$) is created by dissipation of an applied current density
of $\unit[20]{kAcm^{-2}}$. The simulation takes Joule heating due
to current flow and heat dissipation, transport and radiation into
account with a base temperature of $\unit[300]{K}$. Here we have
used a heat conductance $k_{SiO_{2}}=\unit[0.95]{Wm^{-1}K^{-1}}$
for thin SiO$_{2}$ tunnel barriers \cite{Yamane2002}. Taking the
large thermal conductivity of the Au pads and wire bonding into account,
the surface temperature of the contact pads was fixed to $\unit[300]{K}$.
This results in a temperature gradient $\Delta T=\unit[500]{mK}$
between the Si ($T_{Si}$) and Co ($T_{FM}$). Such temperature difference
yields to a nonequilibrium electron distributions in the Co and Si
near the Fermi energy. This results in tunneling of equal number of
electrons in opposite directions, giving rise to a zero net charge
current. However, as the energy dependence of $TSP$ is known to be
different for the forward and reverse tunneling processes \cite{Valenzuela2005,Park2007},
a spin accumulation ($\Delta\mu$) can be induced in the Si (Fig.
1c) \cite{LeBreton2011a,Jansen2012}. 

\begin{figure}[h]
\begin{centering}
\includegraphics{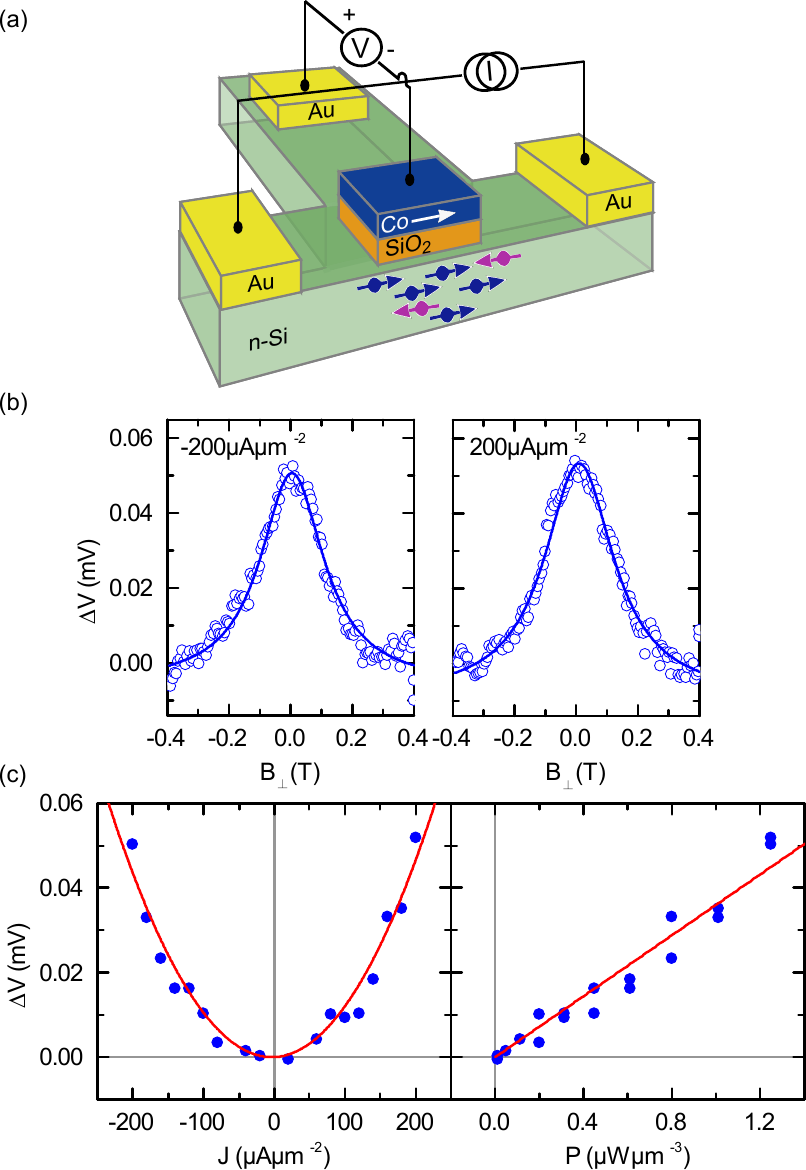} 
\par\end{centering}

\protect\caption{\textbf{(Color online) Thermal Hanle effect at room temperature.}
(a) Device geometry and measurement scheme showing the Si Joule heating
by a DC current , resulting in $T_{Si}>T_{FM}$. The voltage across
the ferromagnetic tunnel contact is measured, with reference to another
Cr/Au contact placed on Si. (b) Thermally induced spin accumulation
in Si, detected as a voltage change ($\Delta V$) by the Hanle effect
with perpendicular magnetic field. The plots show the data for heating
current in two opposite directions ($\pm\unit[200]{\mu A\mu m^{-2}}$)
at $\unit[300]{K}$. The solid lines are Lorentzian fits to the Hanle
curves. (c) Left panel: Thermal Hanle signal $\Delta V$ (symbols)
versus Joule heating current density, with a quadratic fit (solid
line). Right panel: $\Delta V$ (symbols) as a function of Joule heating
power, and a linear fit (solid line). }

\label{fig2-1} 
\end{figure}

Before performing thermal spin injection experiments we have tested
the contacts for electrical spin injection and verified the energy
or bias dependence of $TSP$. The measurements were performed in a
three-terminal Hanle geometry as shown in Fig. 2a. Electrical spin
injection creates a majority spin accumulation and splitting of the
electrochemical potential in the conduction band of the Si. The Hanle
effect is used for the controlled reduction of the induced spin accumulation
by applying a magnetic field ($B_{\bot}$) perpendicular to the FM
magnetization direction. The spin accumulation decays as a function
of $B_{\bot}$ with an approximately Lorentzian line shape given by
, where $\Delta\mu(0)$ and $\Delta\mu(B_{\bot})$ are the spin accumulations
in zero and finite perpendicular magnetic field, respectively, and
$\tau_{sf}$ is the lower limit for spin lifetime \cite{Dash2009,Dankert2013}.
Figure 2b shows the measurement of the electrical Hanle signal for
bias currents of $\pm\unit[250]{nA\mu m^{2}}$ (corresponding bias
voltage of $\pm\unit[300]{mV}$) at $\unit[300]{K}$. The spin injection
($V_{Si}-V_{FM}>0$) from the Co/SiO$_{2}$ tunnel contact creates
a majority spin ($n \uparrow$) accumulation, whereas the spin extraction
($V_{Si}-V_{FM}<0$) creates a minority spin ($n\downarrow$) accumulation
in the Si. As expected the measured Hanle signal for majority and
minority spin accumulation are found to be of opposite sign. The spin
accumulation $\Delta\tau$ scales with the tunnel spin polarization
($TSP$) of the injected current and with spin life time ($\tau_{sf}$).
The detected voltage due to spin accumulation can be written as $\Delta V=TSP\cdot\frac{\Delta\mu}{2}$.
This results in the detected spin signal (spin resistance area product,
spin-RA) which is proportional to $TSP^{2}\cdot\tau_{sf}$. The variation
of $TSP^{2}$ vs. bias voltage is presented in Fig 2d and is found
to be asymmetric. The strong decay of $TSP^{2}$ for energies above
$E_{\text{F}}$ ($V_{Si}-V_{FM}<0$) results in the direction dependent
spin polarized tunnel currents \cite{LeBreton2011a,Jansen2012}. The
difference in spin polarization of the currents tunneling into and
out of the Si can yield to a spin injection in our Co/SiO$_{2}$/n-type
Si devices \cite{LeBreton2011a,Valenzuela2005,Park2007}. This asymmetry
in the $TSP^{2}$ of Co/SiO$_{2}$/n-type Si contact is essential
to induce the large thermal spin accumulation in Si via Seebeck spin
tunneling mechanism. 

\begin{figure}[h]
\begin{centering}
\includegraphics{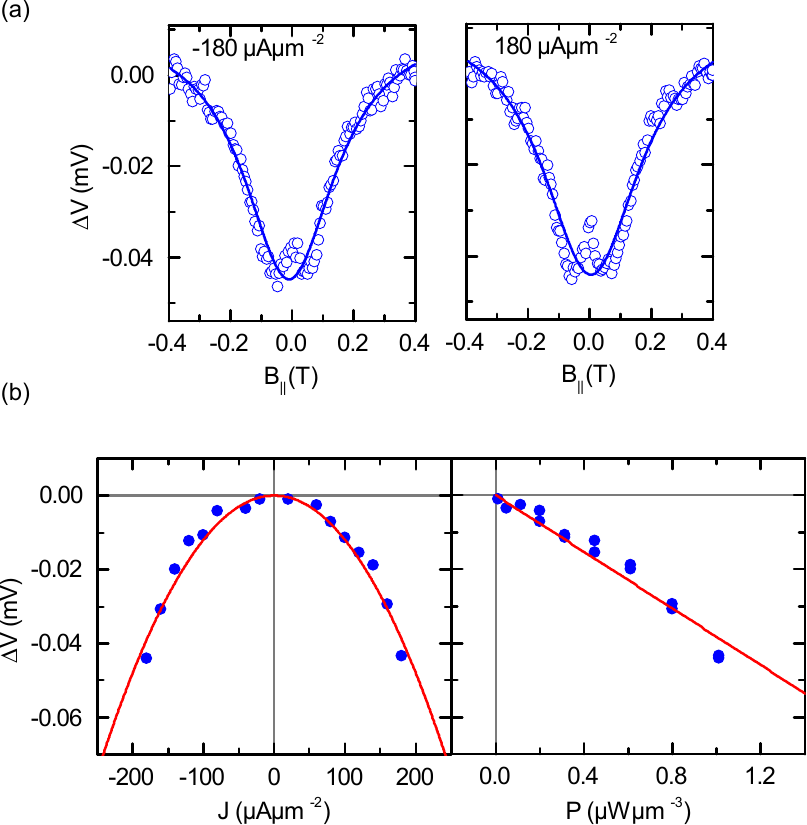} 
\par\end{centering}

\protect\caption{\textbf{(Color online) Thermal inverted Hanle effect at room temperature.}
(a) Thermally induced spin accumulation in Si, detected as a voltage
change ($\Delta V$) employing inverted Hanle effect with in-plane
magnetic field. The plot shows the inverted Hanle data for heating
currents of $\pm\unit[180]{\mu A\mu m^{-2}}$ in two opposite directions.
(b) Left panel: Thermal inverted-Hanle signal $\Delta V$ (symbols)
versus Joule heating current density, together with a quadratic fit
(solid line). Right panel: $\Delta V$ (symbols) as a function of
Joule heating power, and a linear fit (solid line). }

\label{fig2-1-1} 
\end{figure}
\begin{figure*}[t]
\begin{centering}
\includegraphics{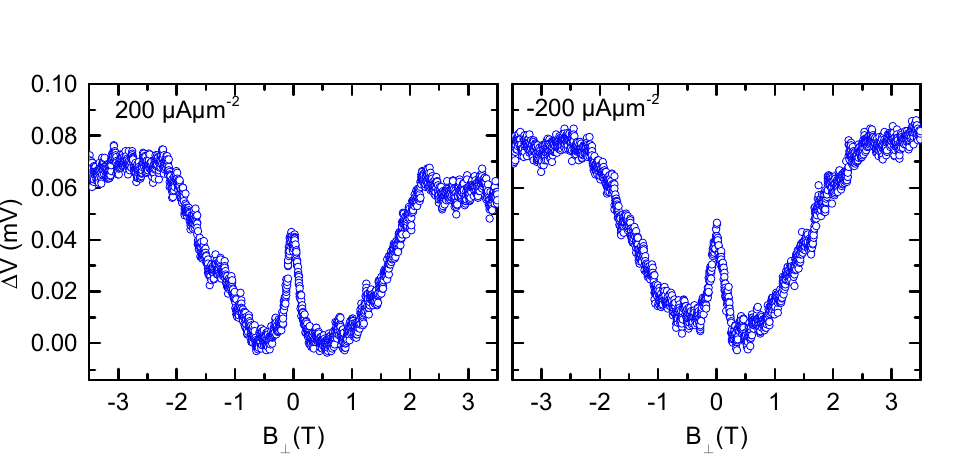} 
\par\end{centering}

\protect\caption{\textbf{(Color online) }Detection of thermally induced spin accumulation
in Si with perpendicular field up to $\unit[3.5]{T}$, for Si heating
currents of $\pm\unit[200]{\mu A\mu m^{-2}}$. It shows the rotation
and saturation of magnetization of ferromagnetic Co electrode to the
out-of-plane direction at higher magnetic fields. }

\label{fig2-1-1-1} 
\end{figure*}
The thermally induced electron spin accumulation in n-type Si has
been created by introducing a temperature gradient between the Si
($T_{Si}$) and Co ($T_{FM}$) separated by a thin SiO$_{2}$ tunnel
barrier ($\unit[1.5]{nm}$). We increase the temperature of Si channel
($T_{Si}$) by a DC current due to Joule heating. The SiO$_{2}$ tunnel
barrier prevents the current and heat from spreading into the ferromagnetic
electrode. The temperature gradient created across the SiO$_{2}$
barrier ($T_{Si}>T_{FM}$) results in spin polarized Seebeck tunnelling
through the oxide barrier, since the $TSP$ for ingoing and outgoing
currents differs. This creates a spin accumulation in the Si conduction
band and therefore a splitting of the electrochemical potential ($\Delta\mu$)
as shown in Fig 1c. The detection of a thermally induced spin accumulation
in Si is performed in both Hanle and inverted Hanle geometry using
a SiO$_{2}$/Co contact \cite{LeBreton2011a,Dash2009}. A voltage
signal is detected while sweeping a magnetic field, under the condition
of zero net tunnel charge current ($I_{tunnel}=0$) between the Co
and the Si \cite{LeBreton2011a,Dash2009}. 

A significant thermal Hanle spin accumulation signal of $\Delta V_{H}=\unit[50]{\mu V}$
is observed in the Si for $T_{Si}>T_{FM}$, for a Joule heating current
density of $J=\pm\unit[200]{\mu A\mu m^{-2}}$ (Fig. 3b). The Lorenzian
line shape of the thermal Hanle signal obtained here is consistent
with our electrical Hanle measurements. The magnitude and sign of
the thermal Hanle curves are also identical for both directions of
the Joule heating current. This is expected, since the Joule heating
and therefore the gradient should be independent of current direction.
For $T_{Si}>T_{FM}$ a spin polarization identical to the electrical
spin injection is observed, which corresponds to majority spin accumulation
in the Si. This should be the case for a decaying $TSP$ above $E_{F}$
as confirmed by our electrical bias dependence measurements presented
in Fig. 2d. Figure 3c shows the amplitude of the thermal Hanle signal
measured for different Joule heating currents. The amplitude of the
thermal spin signal is found to scale quadratically with the heating
current density and hence linearly with the applied heating power,
which confirms the thermal origin of the Hanle signal \cite{LeBreton2011a}. 

Furthermore, we employ the inverted Hanle effect \cite{Dash2011a}
to detect the thermal creation of spin accumulations in Si at room
temperature. The local magnetostatic fields arising from interface
roughness in a ferromagnetic tunnel contact is known to reduce the
spin accumulation in semiconductors. The inverted Hanle effect serves
as experimental signature which recovers the reduced spin accumulation
by applying an in-plane magnetic field $B_{||}$ \cite{Dash2011a}.
Figure 4 shows the measured data due to thermal spin injection for
the case of $T_{Si}>T_{FM}$ in the same Co/SiO$_{2}$/n-type Si device.
An inverted thermal Hanle signal of $\Delta V_{iH}=\unit[45]{\mu V}$
could be detected for a current density of $J=\pm\unit[180]{\mu A\mu m^{-2}}$
(Fig. 4a). As the Joule heating is independent of the current direction,
spin signals of same sign and magnitude are obtained for both current
directions. Figure 4b shows the amplitude of the inverted thermal
Hanle signal for different Joule heating currents, which is also found
to scale quadratically with the heating current density and linearly
with the applied heating power. This scaling of the spin signal is
consistent with a thermally induced spin accumulation in Si \cite{LeBreton2011a}.

The total magnitude of the thermal spin accumulation in n-type Si
is the sum of the Hanle ($\Delta V_{H}$) and inverted Hanle ($\Delta V_{iH}$)
amplitudes \cite{Dash2011a}. A total thermal spin signal of $\Delta V=\unit[95]{\mu V}$
corresponds to a spin splitting $\Delta\mu=2\Delta V/TSP=\unit[0.54]{meV}$
($TSP=0.35$, assumed for SiO$_{2}$/Co contact). This corresponds
to an electron spin polarization of 1\% in the n-type Si conduction
band created by thermal spin injection at room temperature, considering
a parabolic conduction band and a Fermi\textendash Dirac distributions
for each spin direction ($n\uparrow$ and $n\downarrow$). The magnitude
of the thermal spin accumulation defined in terms of a Seebeck spin
tunneling coefficient $S_{st}=\Delta\mu/\Delta T$, is found to be
$\unit[1.08]{meV/K}$. There can be some discrepancy with theory due
to inaccuracy in the determination of $\Delta T$, as experimental
values for the thermal conductance of different materials and their
interfaces are unknown. For the cases of thin tunnel barriers, the
interface thermal resistance dominates over the bulk values and also
differs with sample preparation conditions. In addition, the ballistic
phonon transport and confinement effects in ultra-thin dielectrics
and interface roughness play an important role in altering the thermal
resistance of the contacts \cite{Yamane2002}.

The linear dependence of the spin signal on the heating power confirms
the thermal origin of spin accumulation in n-type Si. Furthermore,
we can rule out other thermomagnetic effects such as spin related
Hall, Nernst, and Ettingshausen effect, because they would not produce
the characteristic Lorentzian shape Hanle signal \cite{LeBreton2011a}.
Nevertheless, to rule out all known thermomagnetic effects, measurements
at high magnetic fields $B_{\bot}$ were performed. For large enough
$B_{\bot}$ the magnetization of the FM rotates out of the plane.
This aligns the injected spins parallel to the magnetization direction
and produces zero precession, resulting in a recovery of the dephased
spin signal. This recovery of the spin signal occurs only if the spin
accumulation is produced by a transfer of the spins from the FM. The
experimental data for the spin accumulation induced by heating of
the Si electrode for $J=\pm\unit[200]{\mu A\mu m^{-2}}$ are shown
in Fig. 5. Indeed, the recovery of the spin accumulation is observed
when the magnetic field increases above $\unit[0.5]{T}$, with the
signal saturating at about $\unit[2.5]{T}$. This corresponds to the
field at which the magnetization of the Co electrode has reached the
full out-of-plane orientation \cite{Sharmaa,Sharma2014}. The observation
of the high-field recovery of the spin accumulation rules out the
spin-Nernst effect and any other mechanism that does not involve transfer
of spins from the FM to the Si \cite{LeBreton2011a}.

In summary, we demonstrate the thermal creation of electron spin polarization
in n-type Si using the Seebeck spin tunneling mechanism at room temperature.
We show that the magnitude of thermally created electron spin polarization
scales quadratic with the Joule heating current density and linearly
with the heating power. A temperature gradient of few hundred milli
kelvins across the ferromagnetic tunnel contacts creates a spin signal
of more than $\unit[95]{\mu V}$, which corresponds to a spin splitting
of $\unit[0.54]{meV}$ and spin polarization of 1\% in the conduction
band of n-type Si. Furthermore, the control experiment with thermal
Hanle measurements at higher magnetic field and the comparison with
electrical spin injection experiments rules out any other spurious
effects. The advantage of utilizing the Seebeck spin tunneling mechanism
is that the magnitude of spin accumulation is not limited to the $TSP$
of ferromagnetic tunnel contact, but can be increased beyond that
due to its dependence on energy derivative of $TSP$. Recently, similar
results using MgO tunnel barrier have been reported, during review
process of our manuscript \cite{Jeon2013}. By carefully designing
the interfaces, Seebeck spin tunneling could be useful for creating
spin polarizations in non-magnetic materials by itself or together
with electrical spin injection, and also for the reusing dissipated
heat in electronic devices and circuits. 

\textbf{Acknowledgement:} The authors acknowledge the support of colleagues
at the Quantum Device Physics Laboratory and Nanofabrication Laboratory
at Chalmers University of Technology. We would also like to acknowledge
the financial supported from the Nano Area of the Advance program
at Chalmers University of Technology, EU FP7 Marie Curie Career Integration
grant and the Swedish Research Council (VR) Young Researchers Grant. 


%

\end{document}